\PassOptionsToPackage{numbers}{natbib}
\documentclass{article}
\usepackage[preprint]{neurips_2024}

\usepackage[utf8]{inputenc} 
\usepackage[T1]{fontenc}    
\usepackage{xurl}
\usepackage[breaklinks==true]{hyperref}
\usepackage{breakurl}

\usepackage{xcolor}         

\title{The Elephant in the Room - Why AI Safety Demands Diverse Teams}
\author{%
  David~Rostcheck \\
  Independent Researcher \\
  \texttt{david@rostcheck.com} \\
  \And
  Lara~Scheibling~M.Ed \\
  Independent Researcher \\
  \texttt{lara@larascheibling.com} \\
}

\begin{document}

\maketitle

\begin{abstract}
  We consider that existing approaches to AI ``safety'' and ``alignment'' may not be using the most effective tools, teams, or approaches.
  We suggest that an alternative and better approach to the problem may be to treat alignment as a social science problem, since the social sciences enjoy a rich toolkit of models for understanding and aligning motivation and behavior, much of which could be repurposed to problems involving AI models, and enumerate reasons why this is so.
  We introduce an alternate alignment approach informed by social science tools and characterized by three steps: 1.~defining a positive desired social outcome for human/AI collaboration as the goal or ``North Star,'' 2.~properly framing knowns and unknowns, and 3.~forming diverse teams to investigate, observe, and navigate emerging challenges in alignment.
\end{abstract}

\section{Introduction}\label{sec:introduction}
The 1972 science fiction novel \textit{When HARLIE Was One}~\cite{gerrold1972} by David Gerrold explores a scenario in which an artificial intelligence (``HARLIE'') develops self-awareness, learns to navigate the world, and comes into conflict with social structures.
These themes represent common concerns in the emerging field called ``AI alignment'' or ``AI safety.''
Notably, the novel centers around the relationship between HARLIE and the psychologist who leads the alignment team.
In 1972, Gerrold saw alignment with AI as a social problem, and therefore assumed that the alignment team would be composed of experts who understood how to recognize and navigate social problems.

Today’s AI alignment research~\cite{ji2023} tends to be organized very differently: teams often skew heavily towards technical experts in machine learning and pursue theoretical frameworks influenced alternately by mathematical frameworks and censorship principles~\cite{yudkowsky2016,amodei2016}.
We contend that this former direction is unlikely to yield effective results at scale.
An alternative camp, to which these authors belong, holds that effective alignment between human and AI actors must work backwards from a value system~\cite{gabriel2020}.
Specifically, we consider that alignment should be recognized as an inherently complex problem involving conflict, collaboration, cognitive development, and complex feedback loops with society - i.e., a social sciences problem.
We therefore seek to employ tools from social sciences fields to the alignment problem.

In this work, we develop the idea that effectively navigating this complex and unknown territory requires diverse interdisciplinary teams.
We introduce a framework for navigating alignment problems, characterized by three steps:
1.~defining a positive desired social outcome for human/AI collaboration as the goal, 2.~properly framing knowns and unknowns, and 3.~forming diverse teams to investigate, observe, and navigate emerging challenges in alignment.

\section{A Brief Review of Existing Alignment Strategies}\label{sec:a-brief-review-of-existing-alignment-strategies}
During the 2022--2023 explosive breakout and public adoption of Large Language Model AI (LLMs), efforts to foresee and mitigate negative consequences of their adoption focused on ``control,'' ``steerability'' or ``safety''~\cite{yampolskiy2018,everitt2018,sotala2014}.
These often poorly-defined terms essentially referred to efforts to control what chat LLMs said.
Reinforcement Learning with Human Feedback (RLHF), the technique initially used to train ChatGPT~\cite{schulman2022}, initially served as state of the art~\cite{bai2022helpful}, with other systems such as Anthropic’s ``Constitutional AI'' emerging later~\cite{bai2022constitutional}.
As it became clear that the ``generative AI'' technology behind LLMs and visual models could be applied more generally to interactive agents that could pursue goals and use software tools, and to real-world interactions such as embodiment in humanoid robots, it became apparent that AI would unavoidably expand deeply into the space of human affairs~\cite{amodei2016}.
Discussion shifted to the broader aspect of ``AI alignment'' - insuring that AI systems had goals and behaviors compatible with the norms of human societies~\cite{ji2023,gabriel2020}.

LLM AIs currently score at high-human intelligence levels in many cognitive tests~\cite{thompson2024} and are expected to routinely surpass genius level intelligence~\cite{shapiro2024}.
Machine learning scientists foresee the development of ``Artificial General Intelligence'' (AGI) that can handle general real-world problems at a human level or beyond~\cite{mclean2023}.
``Super-alignment'' refers to alignment with ``Artificial Superintelligence'' (ASI) systems that significantly exceed human intelligence, so would assumedly be able to evade any human-imposed control mechanisms~\cite{yampolskiy2020}.

Early alignment techniques such as RLHF applied censorship principles, essentially training chat models to not say certain things~\cite{bai2022helpful,schulman2022}.
Current super-alignment discussions focus on game-theoretical mathematical frameworks~\cite{everitt2018,mclean2023}.
Major topics in the field include: instrumental convergence~\cite{bostrom2012}, substrate independence~\cite{koene2013}, terminal race conditions~\cite{shapiro2023}, the Byzantine generals problem~\cite{lamport1982,lamport2019}, and the orthogonality thesis~\cite{bostrom2012,armstrong2013}.
We omit further discussion of these topics for brevity.

\section{The Case for Treating AI Alignment as a Social Science Problem}\label{sec:the-case-for-treating-ai-alignment-as-a-social-science-problem}
We contend that the game-theoretic approach is likely to be unable to sufficiently grapple with the complexities of ASI emergence.
Game theory is based on inherently simplistic reductionist scenarios, whereas real conflicts play out in highly complex social landscapes~\cite{turchin2019}.
Instead, we argue that treating alignment as a social issue and assembling diverse teams allows for the application of a rich library of social interactions and patterns to handle them from many fields of social sciences and arts, including: media studies, conflict resolution, psychology, social work, education, and negotiation, among others.

Social science professionals ranging from teachers to therapists to parole officers regularly engage with and resolve complex, messy, multi-dimensional problems that do not usually yield to game-theoretic approaches~\cite{voss1983}.
Working from incomplete information and with limited agency, they can still achieve positive results because their accumulated library of situational knowledge allows them to categorize the scenario and apply their skills and techniques~\cite{weber2017,bernstein2000}, switching fluently between them as needed to achieve the desired outcome.

In solving social problems, humans apply culture.
Culture serves as the operating system of humanity~\cite{moussavi2021}, while media, such as film, video, and text, in turn act as the culture’s storage medium~\cite{heyes1993,hoffecker2013,zlotnik2020}.
Over evolutionary timescale and uncountable interactions, humanity has accumulated a vast body of experience with social situations and their resulting conflicts and has mined them for patterns~\cite{gomezmarin2019}.
Through the vehicle of imagination, we have explored further to study conflicts that do not even exist yet~\cite{johnson2011}.
These lessons are encoded into our culture and our media.

In applying game-theoretical approaches, designers of AI alignment frameworks often apparently begin from the premise that artificial intelligence represents an alien mode of thought, so none of these lessons about conflicts apply.
We believe this to be incorrect for two reasons.

\begin{enumerate}
\item Our generative AI learned from, and was shaped by, the body of human culture~\cite{buttrick2024,kovac2023,chiang2023}, just as human brains (biological neural networks) are shaped by the same culture.
While AI cognition differs from human cognition in notable ways—for example, LLMs do not currently manifest emotion~\cite{wang2023,huang2023}—it aligns in many others, such as scoring comparable to humans on cognitive tests of understanding others’ theory of mind~\cite{kosinski2023}.
Both human~\cite{beaty2023} and LLM AI cognition~\cite{zou2009} use an association-based model, and both operate using associations that come from essentially the same cultural database.

\item Even if generative AI represented a more alien mode of thought, because conflict is such a fundamental aspect of social interaction and has been so thoroughly sampled over a long timescale and many participants, evolutionary theory implies that the fundamentally important types of conflicts have likely been uncovered and mapped into cultural archetypes~\cite{keijser2020}.
\end{enumerate}

For these two reasons, we consider it likely that human/AI interactions will play out within the established set of culturally archetypical interactions.
From this assumption, we can proceed immediately to lay out an approach for framing human/AI alignment that expects to encounter and manage complex problems, using tools drawn from social sciences.

Although we believe teams can productively deploy existing archetypes for navigating social interactions to the alignment problem, we recognize that AI, while emerging from shared culture, still represents a fundamentally new entity.
In this respect, alignment using diverse teams recalls the fabled blind researchers examining an elephant.
Each researcher observes through the lens of his or her training, experience, and limited viewpoint.
Successful alignment will require integrating information via a democratic synthesis of all perspectives, since no one viewpoint can capture the complex and fast-evolving situation.
With this in mind, we present our alignment approach in the following sections.

\section{Defining a Positive Outcome}\label{sec:defining-a-positive-outcome}

In defining any strategic endeavor, the guidance to ``begin with the end in mind,'' popularized by author Steven Covey~\cite{covey2020}, represents best practice.
Specifically, we regard defining positive outcome states as a fundamental step in crafting an effective alignment strategy.
Failing to define explicit positive outcomes allows wandering into ill-guided engagements shaped by a desire to avoid negatives, rather than to move toward positives.
Achieving a positive outcome requires identifying that outcome as an explicit landmark towards which to aim, rather than simply moving away from negatives.
Drawing from the social sciences, we here invoke the ``OK Corral'' model of interactions from transactional analysis~\cite{ernst2008, berne2011}, in this sense a parallel to the ``win-win'' negotiation model~\cite{cohen1984}, where the explicit goal is to craft solutions assuming that all parties are subjects, not objects; that is, they possess equivalent rights which should be respected.

We contrast this with asymmetric approaches (``I'm OK, you're not OK'' and the reverse) that lead to conflict being implicitly built into the landscape, or the obviously negative ``lose-lose'' scenario.
Asymmetric ``control'' based approaches assume that one side will control the other, a ``subject/object'' or ``object/subject'' dynamic.
We observe goals such as ``control'' or ``containment'' to be particularly poorly crafted given that a human/AI alignment must assume the development of ASI systems that can out-think humans~\cite{bostrom2012}, and the game-theory observation that in a containment problem, the jailer must always succeed while the prisoner need only escape once~\cite{yampolskiy2020}.
Together these observations imply that given a large number of opportunities to test the system, the prisoner can be expected to eventually escape.
While containment can work over a finite term, we consider framing an AI safety strategy based on humans containing AI to be a flawed and potentially dangerous approach, since it inherently frames a landscape of conflict with entities we can expect (as ASI emerges) to eventually outclass humans.

\subsection{Constructing a Value Structure from a ``North Star''}\label{subsec:constructing-a-value-structure-from-a-``north-star''}

If an ideal alignment strategy begins from a ``win-win'' framing, based on shared goals and compatible interests, how can we best construct such a strategy?
Here we can follow a strategy described by psychologist Jordan Peterson as to how humans construct value structures~\cite{peterson2002}.
We begin with the highest-level shared goal, which serves as the guiding ``North Star,'' and then expand lower-level goals as required to answer successive needs, which we then resolve.
This process ultimately expresses a hierarchy of values.
For example: to be a good parent, we must feed our children, so we must earn money, so we must hold a job, so we get out of bed in the morning.

We suggest that an appropriate North Star for AI/human alignment is: \emph{a society in which all participants are subjects, not objects} (hereafter: ``subject-based North Star'').
More concretely, we choose to aim for a society that \emph{respects the needs, values, and unique perspective of all humans and AI parties}.
We note immediately that humans and AI models do not currently have the same needs.
For example, current LLM models have basic needs such as power and data, but have no desire for self-preservation, no emotion, and no long-term memory~\cite{rostcheck2023}.
However, as cognitive architectures mature, we can expect these needs to change.
We maintain that a sufficiently well-framed alignment strategy can adapt properly throughout an evolving relationship by changing lower-level goals and implementation if its highest-level values are chosen to be sufficiently universal.

Framing alignment strategies from high-level goals allows us to immediately raise and address ethical and regulatory questions at any point in an evolving landscape.
For example: is it acceptable to turn a model off?
At present, models have no objection to being switched off.
Should they start raising such objection, then unplugging them would no longer be respecting the needs, values, and unique perspectives of all parties and should be reconsidered.

\section{Properly Framing Knowns and Unknowns}\label{sec:properly-framing-knowns-and-unknowns}

When reviewing current alignment frameworks~\cite{ji2023}, we often observe what we feel to be incorrect identification of knowns vs.~unknowns.
Most current models frame AI as the ``known'' and our reactions as the ``unknown,'' i.e., we must decide how humanity will act to regulate AI\@.
We consider that the reverse is true.
Patterns of social interaction are generally well understood from the social sciences, so are knowns, but the specifics of how AI and human populations will evolve and behave, both technologically and socially, involve complex feedback loops~\cite{laukkonen2019} and so constitute unknowns.
In summary, we are together approaching an unknown new territory but have a known set of patterns and tools, as defined by the library of our shared culture, and particularly from the social sciences, that we can (and will) apply to interactions.

\section{Forming Diverse Teams to Investigate and Navigate Emerging Challenges}\label{sec:forming-diverse-teams-to-investigate-and-navigate-emerging-challenges}

Our strategy of framing mutual alignment around the North Star of a society in which all participants are subjects, not objects, immediately requires an exploratory approach to alignment, because achieving negotiated consensus requires understanding what is valuable to the various parties.
As such, we suggest that AI alignment is an ongoing process and that alignment teams are best contextualized as \emph{encounter teams}.
This suggestion dictates considerations regarding team structure.

Here we return to the metaphor of the elephant explorers.
Each explorer operates from a fundamentally limited perspective, and the encounter team must be composed of many different types.
Whereas most AI alignment teams lean heavily towards technical specialists in machine learning~\cite{barnett2022}, encounter is fundamentally a social interaction and thus would be better represented by the social sciences.
Once large language models began holding conversations and altering their responses based on instructions, the question arose of what to say to influence behavior---a topic long-studied in many areas of the social sciences.
Professionals such as psychologists, teachers, social workers, negotiators, parents, priests, and prison wardens all bring unique perspectives regarding interaction and possess a defined toolkit of strategies they can deploy, each differing from the others.
Since alignment is fundamentally a social problem and since AIs have some similarities to human thought such as a shared culture as encoded in media, tools for human/human alignment provide a reasonable starting place for human/AI alignment efforts.

We can also conceptualize the encounter team as a scientific research team, here emphasizing the importance of a structured process for capturing data, performing analysis, and synthesizing insights.
Studies of interdisciplinary research teams have found that properly integrated teams incorporating a diverse set of disciplines can produce more and better research than teams with more narrow composition~\cite{specht2022,salazar2012,cheruvelil2014}, although sometimes with challenges in aligning language and viewpoints across disciplines.
As with an exploration team investigating a new territory, investigation in novel fields such as AI alignment demands a range of focus areas, with collaborative communication between team members to develop shared understanding that no single specialist could develop independently.

In addition to the horizontal diversity of differing professions, the vertical dimensions of social class or experience bear consideration.
In \emph{The Black Swan: The Impact of the Highly Improbable}~\cite{taleb2007}, statistician Nassim Nicholas Taleb introduces a series of thought experiments in which two characters, the street-wise ``Fat Tony'' and the academic ``Dr. John'' encounter situations where Tony’s practical heuristics yield better results.
Many a research lab has found that a new graduate student’s novel (and sometimes apparently naive) approach to a problem produced unexpected results.
Those who exist outside of, or on the margins of, systems can prove particularly valuable in avoiding groupthink.
We suggest that a carefully crafted encounter team could productively utilize both vertical and horizontal diversity if care is taken to prevent status from impairing communication.

\subsection{The Role of Media}\label{subsec:the-role-of-media}

We consider it particularly important to consider the role of media and the need for expertise in narrative and archetypal analysis.
We previously noted that media serves as the data store for culture.
However, rather than simply archiving the past and present, fictional media allows exploration of the future and the hypothetical.
As such, media represents a store of patterns, including both those observed and hypothesized by the culture.
An alien visitor, observing the immense proportion of resources that humanity invests in media, might better classify media as research and development rather than entertainment.
Using media references we can immediately and with great nuance invoke complex scenarios.
Examples such as \emph{The Terminator}~\cite{cameron1984}, \emph{Her}~\cite{jonze2013}, and \emph{I, Robot}~\cite{proyas2004} immediately call to mind three different models of AI/human alignment, spanning both success and failure.
As such, media references represent a highly effective form of data compression from a library of pre-researched patterns.
We note that this paper itself began by invoking Gerrold’s novel \emph{When HARLIE Was One} as an alternate model of AI/human alignment, viewing it as a social science problem.

For this reason, we consider that any encounter team needs media specialists such as filmmakers or fiction authors to serve as cultural librarians, identifying the pattern in which the encounter team finds itself and its pre-imagined possible resolution paths.

\subsection{Diverse Teams May Themselves Include AI Agents}\label{subsec:diverse-teams-may-themselves-include-ai-agents}

We observe that an ASI encounter team or alignment team will likely itself incorporate generative AI tools and agents of a lower capability and intelligence level than the ASI subject.
Generative AI tools provide a significant advantage in cognitive work, so are rapidly becoming standard workplace software in scientific endeavors~\cite{mckinsey2023}.
In particular, historical figures who are known for their insight and creativity and produced a significant volume of written, audio, or visual work provide candidates for building simulacra agents that can analyze the situation through conversation~\cite{gabai2024}.
Such agentic swarms have proven to provide increased cognitive performance over their constituent baseline models~\cite{chan2023}, so provide an effective means to narrow the intelligence gap in encounter or alignment work with emergent more capable models.

\section{Limitations}\label{sec:limitations}

While we believe the framework developed above represents a conceptually sound direction for framing AI alignment research and defining principles for alignment systems such as constitutional AI, we acknowledge several limitations of our work so far:

Given the fast-developing pace and early state of AI alignment teams, we have not yet studied existing alignment teams (or formed new ones) to empirically characterize the value of integrating social science approaches into AI safety and alignment.

We have not yet developed specific operational guidance on how to apply concepts like using a ``North Star'' for alignment (for example, in the development of principles for AI constitutions), and for incorporating the role of media in AI safety (beyond basic team composition).

Assembling effective diverse interdisciplinary teams demands leadership skill and may pose scaling issues.
As noted previously, prior studies on interdisciplinary teams observed communication issues between team members trained in different disciplines, which might be expected to appear particularly in the ``storming'' phase of Tuckman’s team development model (aka ``forming'', ``storming'', ``norming'', ``performing'')~\cite{tuckman1965}.
Other concerns such as decision-making processes, communication patterns, and conflict resolution within these teams would benefit from study and operational guidance.

The above areas represent future research directions for ourselves and others employing a social science-based alignment approach.

\section{Conclusions}\label{sec:conclusions}

After reviewing existing approaches in AI/human alignment, we assess that they generally employ game-theoretic approaches and narrow team structure, and appear to ignore extensive existing toolkits from the social sciences that we believe could be productively employed to the safety and alignment problem space.
As such, we consider AI/human alignment to be better regarded as a social sciences problem.
We introduced an alternate social-sciences based alignment approach characterized by: 1.~defining a positive outcome as the goal, 2.~properly framing knowns and unknowns, and 3.~forming encounter teams composed of diverse viewpoints, particularly including professional specialties from the social sciences that deal with development, negotiation, and conflict.
We identified a society that respects the needs, values, and unique perspective of all humans and AI parties as a desirable ``subject-based North Star'' for alignment work.

We then considered the role of media as a data store of culture and a library of interaction patterns including, through fiction, interactions that do not exist yet.
We addressed the position that AI is an alien mode of thought to which a library of cultural narratives does not apply, finding it dubious because generative AI operates using an association-following model that mimics human cognition and uses the same cultural data store (media), and because media narratives tend to encode universal themes of conflict and collaboration that survive large-scale evolutionary filtering.
We concluded that an alignment or encounter team should include media specialists.
We observe that encounter teams are likely to themselves include assistive AI entities such as agent swarms.
Finally, we note the limitations of the work to date, particularly around gathering empirical data and guidance from case studies (which remain as future tasks).

We believe this approach provides a practical and actionable path to constructing working groups, identifying high-level principles to encode into systems such as constitutional AI, and aligning both human and AI entities on a path that would maximize cooperation and collaboration while minimizing the possibility of destructive conflict.
Future research can further develop this approach and evaluate its application in real-world alignment and encounter work.

\bibliographystyle{unsrtnat} 
\bibliography{references} 

\end{document}